\documentclass[article,12pt]{article}  
\renewcommand{\baselinestretch}{1.0} 
\usepackage{geometry}			
\usepackage{bm}
\usepackage{verbatim}
\usepackage{pdfsync}
\usepackage{times}

\usepackage[usenames]{color}

\newcommand{\MyCommands}{
\setlength{\textheight}{8.5in}
\setlength{\topmargin}{-1cm}
\setlength{\textwidth}{6in}
\newcommand{\card}[1]{\mid\!\mid\!{#1}\!\mid\!\mid}
\newcommand{\stt}[1]{\small\tt #1}
\newcommand{\cross}{\ \mbox{\large $\times$}\ }
\newcommand{\lbrac}{[\![}
\newcommand{\rbrac}{]\!]}
\newcommand{\meet}{\wedge}
\newcommand{\join}{\vee}
\newcommand{\below}{\sqsubseteq}
\newcommand{\waybelow}{\ll}
\newcommand{\MEET}{\bigwedge}
\newcommand{\JOIN}{\bigvee}
\newcommand{\ggets}{\mbox{:= }}
\newcommand{\entails}{\models}
\newcommand{\restrict}[1]{\!\upharpoonleft_{#1}}
\newcommand{\QEE}{\hfill\triangle}
\newcommand{\hhdownarrow}{\makebox[0mm]{\raisebox{0.3ex}[0mm][0mm]{$\downarrow$}}\makebox[0mm]{$\downarrow$}\,\,}
\newcommand{\low}[1]{\raisebox{-0.3ex}{#1}}
\newcommand{\PIII}{\Pi_{1}^{1}}
\newcommand{\SIII}{\sum_{1}^{1}}
\newcommand{\CK}{\omega_{1}^{\mbox{ck}}}
\newcommand{\lh}{<_{H}}
\newcommand{\leh}{\leq_{H}}
\newcommand{\makeclause}[3]%
  {#1\ \leftarrow\ #2\ \&\ \ldots\ \&\ #3}
\newcommand{\memb}{\ \epsilon\ }
\newcommand{\ms}[1]{\mbox{\stt{#1}}}
\newcommand{\Iff}{\Longleftrightarrow}
\newcommand{\skp}[1]{\vspace{#1\baselineskip}}
\newcommand{\dispform}[2]%
      {\renewcommand{\theequation}{#1}
       \begin{equation}
       #2
       \end{equation}
      }
\newcounter{equn}[section]
\newcommand{\eqnsect}[1]%
      {\refstepcounter{equn}
       \renewcommand{\theequation}{\thesection.\theequn}
       \begin{equation}
       #1
       \end{equation}
      }
\newcommand{\Nat}{\mathbb{N}}
\newcommand{\dos}{$\ldots\ $}
\newcommand{\Reals}{\hspace{0.2em}R\hspace{-0.9em}I\hspace{0.5em}}
\newcommand{\TPJ}[2]{\mbox{{\bf T}}_{#1}^{#2}}
\newcommand{\TQJ}{\mbox{{\bf T}}_{Q}^{\mbox{J}}}
\newcommand{\TQ}[1]{\mbox{{\bf T}}_{#1}}
\newcommand{\TP}{\mbox{{\bf T}}_{P}}
\newcommand{\abs}[1]{\mid#1\mid}
\newcommand{\abss}[1]{\mid\mid#1\mid\mid}
\newcommand{\HB}[1]{\mbox{B}^{#1}_{{\mbox{{\bf L}}}}}
\newcommand{\HBN}[2]{\mbox{B}^{#1}_{#2}}
\newcommand{\PHB}[1]{2^{\mbox{B}_{#1}}}
\newcommand{\QED}{\rule{2mm}{3mm}}
\newcommand{\QEDD}{\hspace*{\fill}\rule[-2pt]{2mm}{3mm}}
\newcommand{\OPGAM}{\Gamma\ :\ (P\omega)^{n} \longrightarrow\ P\omega}
\newcommand{\re}{\sum_{1}^{0}}
\newcommand{\ARGS}%
    {\mbox{lub}\ \pi_{1}[V],\ \ldots,\ \mbox{lub}\ \pi_{n}[V]}
\newcommand{\SEG}[2]{#1,\ \ldots,\ #2}
\newcommand{\PO}{<\PHB{\mbox{J}},\subseteq>}
\newcommand{\LAF}{L_{{\cal F}}}
\newcommand{\LAPA}{L_{{\cal P}}}
\newcommand{\LAP}{L_{{\cal P}}}
\newcommand{\TUW}[1]{\mbox{{\bf T}}_{#1}\!\!\uparrow\!\!\omega}
\newcommand{\TDm}[1]{\mbox{{\bf T}}_{#1} \downarrow m}
\newcommand{\TDW}[1]{\mbox{{\bf T}}_{#1}\!\!\downarrow\!\!\omega}
\newcommand{\TDWJ}[2]
                    {\mbox{{\bf T}}_{#2}^{#1}\!\!\downarrow\!\!\omega}
\newcommand{\TDNJ}[3]{\mbox{{\bf T}}_{#2}^{#1} \downarrow #3}
\newcommand{\gfp}[1]{\mbox{gfp}(\mbox{\bf T}_{#1})}
\newcommand{\gfpJ}[2]{\mbox{gfp}(\mbox{{\bf T}}_{#2}^{#1})}
\newcommand{\lfp}[1]{\mbox{lfp}(\mbox{{\bf T}}_{#1})}
\newcommand{\Str}{\mbox{{\bf Str}}_{{\cal G}_{0}}}
\newcommand{\RStr}{\mbox{{\bf Str}}_{{\cal G}_{0},\Re}}
\newcommand{\pni}{\par \skp{1} \noindent}
\newcommand{\pnt}{$\bullet$}
\newcommand{\spc}{$\ $}
\newcommand{\absv}[1]{\mid\!#1\!\mid}
\newcommand{\guard}{\raisebox{0.3ex}
                             {\framebox[0.05in]{\rule{0mm}{0.03in}}}}
\newcommand{\und}[1]{\underline{#1}}
\newcommand{\ClEq}{\mbox{ClEq}}
\newcommand{\seb}{[\,{\cal A}\,]_{\Sigma}} 
\newcommand{\lb}{\mbox{
\begin{picture}(0,8)(0,0)
\put(0,5){\line(0,3){3}}
\put(0,8){\line(1,0){3}}
\end{picture}         }
                }
\newcommand{\rb}{\!\mbox{
\begin{picture}(0,8)(0,0)
\put(3,5){\line(0,3){3}}
\put(0,8){\line(1,0){3}}
\end{picture}         }
                }
\newcommand{\nm}{\mbox{{\bf i}}}
\newcommand{\domp}{\mid\!\!{\cal M}'\!\!\mid}
\newcommand{\dompp}{\mid\!\!{{\cal M}'}'\!\!\mid}
\newcommand{\lang}{\mbox{{\bf L}}}
\newcommand{\dotcup}{\ensuremath{\mathaccent\cdot\cup}}
\renewcommand{\baselinestretch}{1.0}

\newtheorem{Experiment}{Experiment}[section]
\newtheorem{NS-Experiment}{Experiment}

\newtheorem{Fact}{Fact}[section]
\newtheorem{NS-Fact}{Fact}

\newtheorem{Example}{Example}[section]
\newtheorem{NS-Example}{Example}

\newtheorem{Definition}{Definition}[section]
\newtheorem{NS-Definition}{Definition}

\newtheorem{Proposition}{Proposition}[section]
\newtheorem{NS-Proposition}{Proposition}

\newtheorem{Homework}{Homework}[section]
\newtheorem{NS-Homework}{Homework}

\newtheorem{Theorem}{Theorem}[section]
\newtheorem{NS-Theorem}{Theorem}

\newtheorem{Observation}{Observation}[section]
\newtheorem{NS-Observation}{Observation}

\newtheorem{Lemma}{Lemma}[section]
\newtheorem{NS-Lemma}{Lemma}

\newtheorem{Remark}{Remark}[section]
\newtheorem{NS-Remark}{Remark}

\newtheorem{Algorithm}{Algorithm}[section]
\newtheorem{NS-Algorithm}{Algorithm}

\newtheorem{Corollary}{Corollary}[section]
\newtheorem{NS-Corollary}{Corollary}

\newtheorem{Claim}{Claim}[section]
\newtheorem{NS-Claim}{Claim}

\newtheorem{NS-claim}{claim}

\newtheorem{Directive}{Directive}[section]
\newtheorem{NS-Directive}{Directive}

\newtheorem{Problem}{Problem}[section]
\newtheorem{NS-Problem}{Problem}

\newtheorem{Solution}{Solution}[section]
\newtheorem{NS-Solution}{Solution}

\newtheorem{Notation}{Notation}[section]
\newtheorem{NS-Notation}{Notation}

\newtheorem{Note}{Note}[section]
\newtheorem{NS-Note}{Note}

\newtheorem{Question}{Question}[section]
\newtheorem{NS-Question}{Question}

\newtheorem{Answer}{Answer}[section]
\newtheorem{NS-Answer}{Answer}

\newtheorem{Counterexample}{Counterexample}[section]
\newtheorem{NS-Counterexample}{Counterexample}

\newcommand{\sect}[1]{\stepcounter{section}$\!\!$
                      {\large {\bf \S\thesection. \spc #1}}
                     }
\newcommand{\follows}{$\!\!\!$\textbf{\em :}\spc\spc}
\newcommand{\ket}[1]{| #1 \rangle}
\newcommand{\bra}[1]{\langle #1 |}
\newcommand{\dirac}[2]{\langle #1 | #2 \rangle}
\newcommand{\st}\backepsilon
\newcommand{\suchthat}{\mathrel{\ooalign{$\ni$\cr\kern-1pt$-$\kern-6.5pt$-$}}}
}
\newcommand{\card}[1]{\mid\!\mid\!{#1}\!\mid\!\mid}
\newcommand{\stt}[1]{\small\tt #1}
\newcommand{\cross}{\ \mbox{\large $\times$}\ }
\newcommand{\lbrac}{[\![}
\newcommand{\rbrac}{]\!]}
\newcommand{\meet}{\wedge}
\newcommand{\join}{\vee}
\newcommand{\below}{\sqsubseteq}
\newcommand{\waybelow}{\ll}
\newcommand{\MEET}{\bigwedge}
\newcommand{\JOIN}{\bigvee}
\newcommand{\ggets}{\mbox{:= }}
\newcommand{\entails}{\models}
\newcommand{\restrict}[1]{\!\upharpoonleft_{#1}}
\newcommand{\QEE}{\hfill\triangle}
\newcommand{\hhdownarrow}{\makebox[0mm]{\raisebox{0.3ex}[0mm][0mm]{$\downarrow$}}\makebox[0mm]{$\downarrow$}\,\,}
\newcommand{\low}[1]{\raisebox{-0.3ex}{#1}}
\newcommand{\PIII}{\Pi_{1}^{1}}
\newcommand{\SIII}{\sum_{1}^{1}}
\newcommand{\CK}{\omega_{1}^{\mbox{ck}}}
\newcommand{\lh}{<_{H}}
\newcommand{\leh}{\leq_{H}}
\newcommand{\makeclause}[3]%
  {#1\ \leftarrow\ #2\ \&\ \ldots\ \&\ #3}
\newcommand{\memb}{\ \epsilon\ }
\newcommand{\ms}[1]{\mbox{\stt{#1}}}
\newcommand{\Iff}{\Longleftrightarrow}
\newcommand{\skp}[1]{\vspace{#1\baselineskip}}
\newcommand{\dispform}[2]%
      {\renewcommand{\theequation}{#1}
       \begin{equation}
       #2
       \end{equation}
      }
\newcounter{equn}[section]
\newcommand{\eqnsect}[1]%
      {\refstepcounter{equn}
       \renewcommand{\theequation}{\thesection.\theequn}
       \begin{equation}
       #1
       \end{equation}
      }
\newcommand{\Nat}{\mathbb{N}}
\newcommand{\dos}{$\ldots\ $}
\newcommand{\Reals}{\hspace{0.2em}R\hspace{-0.9em}I\hspace{0.5em}}
\newcommand{\TPJ}[2]{\mbox{{\bf T}}_{#1}^{#2}}
\newcommand{\TQJ}{\mbox{{\bf T}}_{Q}^{\mbox{J}}}
\newcommand{\TQ}[1]{\mbox{{\bf T}}_{#1}}
\newcommand{\TP}{\mbox{{\bf T}}_{P}}
\newcommand{\abs}[1]{\mid#1\mid}
\newcommand{\abss}[1]{\mid\mid#1\mid\mid}
\newcommand{\HB}[1]{\mbox{B}^{#1}_{{\mbox{{\bf L}}}}}
\newcommand{\HBN}[2]{\mbox{B}^{#1}_{#2}}
\newcommand{\PHB}[1]{2^{\mbox{B}_{#1}}}
\newcommand{\QED}{\rule{2mm}{3mm}}
\newcommand{\QEDD}{\hspace*{\fill}\rule[-2pt]{2mm}{3mm}}
\newcommand{\OPGAM}{\Gamma\ :\ (P\omega)^{n} \longrightarrow\ P\omega}
\newcommand{\re}{\sum_{1}^{0}}
\newcommand{\ARGS}%
    {\mbox{lub}\ \pi_{1}[V],\ \ldots,\ \mbox{lub}\ \pi_{n}[V]}
\newcommand{\SEG}[2]{#1,\ \ldots,\ #2}
\newcommand{\PO}{<\PHB{\mbox{J}},\subseteq>}
\newcommand{\LAF}{L_{{\cal F}}}
\newcommand{\LAPA}{L_{{\cal P}}}
\newcommand{\LAP}{L_{{\cal P}}}
\newcommand{\TUW}[1]{\mbox{{\bf T}}_{#1}\!\!\uparrow\!\!\omega}
\newcommand{\TDm}[1]{\mbox{{\bf T}}_{#1} \downarrow m}
\newcommand{\TDW}[1]{\mbox{{\bf T}}_{#1}\!\!\downarrow\!\!\omega}
\newcommand{\TDWJ}[2]
                    {\mbox{{\bf T}}_{#2}^{#1}\!\!\downarrow\!\!\omega}
\newcommand{\TDNJ}[3]{\mbox{{\bf T}}_{#2}^{#1} \downarrow #3}
\newcommand{\gfp}[1]{\mbox{gfp}(\mbox{\bf T}_{#1})}
\newcommand{\gfpJ}[2]{\mbox{gfp}(\mbox{{\bf T}}_{#2}^{#1})}
\newcommand{\lfp}[1]{\mbox{lfp}(\mbox{{\bf T}}_{#1})}
\newcommand{\Str}{\mbox{{\bf Str}}_{{\cal G}_{0}}}
\newcommand{\RStr}{\mbox{{\bf Str}}_{{\cal G}_{0},\Re}}
\newcommand{\pni}{\par \skp{1} \noindent}
\newcommand{\pnt}{$\bullet$}
\newcommand{\spc}{$\ $}
\newcommand{\absv}[1]{\mid\!#1\!\mid}
\newcommand{\guard}{\raisebox{0.3ex}
                             {\framebox[0.05in]{\rule{0mm}{0.03in}}}}
\newcommand{\und}[1]{\underline{#1}}
\newcommand{\ClEq}{\mbox{ClEq}}
\newcommand{\seb}{[\,{\cal A}\,]_{\Sigma}} 
\newcommand{\lb}{\mbox{
\begin{picture}(0,8)(0,0)
\put(0,5){\line(0,3){3}}
\put(0,8){\line(1,0){3}}
\end{picture}         }
                }
\newcommand{\rb}{\!\mbox{
\begin{picture}(0,8)(0,0)
\put(3,5){\line(0,3){3}}
\put(0,8){\line(1,0){3}}
\end{picture}         }
                }
\newcommand{\nm}{\mbox{{\bf i}}}
\newcommand{\domp}{\mid\!\!{\cal M}'\!\!\mid}
\newcommand{\dompp}{\mid\!\!{{\cal M}'}'\!\!\mid}
\newcommand{\lang}{\mbox{{\bf L}}}
\newcommand{\dotcup}{\ensuremath{\mathaccent\cdot\cup}}
\renewcommand{\baselinestretch}{1.0}

\newtheorem{NS-Experiment}{Experiment}

\newtheorem{NS-Fact}{Fact}

\newtheorem{Example}{Example}[section]
\newtheorem{NS-Example}{Example}

\newtheorem{NS-Definition}{Definition}

\newtheorem{NS-Proposition}{Proposition}

\newtheorem{NS-Homework}{Homework}

\newtheorem{NS-Theorem}{Theorem}

\newtheorem{NS-Observation}{Observation}

\newtheorem{Lemma}{Lemma}[section]
\newtheorem{NS-Lemma}{Lemma}

\newtheorem{NS-Remark}{Remark}

\newtheorem{NS-Algorithm}{Algorithm}

\newtheorem{Corollary}{Corollary}[section]
\newtheorem{NS-Corollary}{Corollary}

\newtheorem{NS-Claim}{Claim}

\newtheorem{NS-claim}{claim}

\newtheorem{NS-Directive}{Directive}

\newtheorem{NS-Problem}{Problem}

\newtheorem{NS-Solution}{Solution}

\newtheorem{NS-Notation}{Notation}

\newtheorem{NS-Note}{Note}

\newtheorem{NS-Question}{Question}

\newtheorem{NS-Answer}{Answer}

\newtheorem{NS-Counterexample}{Counterexample}

\newcommand{\sect}[1]{\stepcounter{section}$\!\!$
                      {\large {\bf \S\thesection. \spc #1}}
                     }
\newcommand{\follows}{$\!\!\!$\textbf{\em :}\spc\spc}
\newcommand{\ket}[1]{| #1 \rangle}
\newcommand{\bra}[1]{\langle #1 |}
\newcommand{\dirac}[2]{\langle #1 | #2 \rangle}
\newcommand{\st}\backepsilon
\newcommand{\suchthat}{\mathrel{\ooalign{$\ni$\cr\kern-1pt$-$\kern-6.5pt$-$}}}

\renewcommand{\thesection}{\arabic{section}}

\usepackage{amsmath}
\usepackage{amsfonts,amssymb}
\usepackage{graphicx}
\usepackage[colorlinks=true, allcolors=black]{hyperref}
\usepackage{parskip}
\usepackage{color}
\usepackage{tikz-cd}
\usepackage{verbatim}
\usepackage{mathtools}
\usepackage{array}

\newcolumntype{M}[1]{>{\hbox to #1\bgroup\hss$}l<{$\egroup}}

\makeatletter
\newcommand\@brcolwidth{0.67em}
\newenvironment{brmatrix}{%
    \left[%
    \hskip-\arraycolsep
    \new@ifnextchar[\@brarray{\@brarray[\@brcolwidth]}%
}{%
    \endarray
    \hskip -\arraycolsep
    \right]%
}
\def\@brarray[#1]{\array{r*\c@MaxMatrixCols {M{#1}}}}
\makeatother

\title{\large\bf Interaction of Multiple Tensor Product Operators of the Same Type: an Introduction}
\author{\normalsize Howard~A.~Blair$^{1,2,3}$ \and H~Shelton~Jacinto$^1$ \and Paul~M.~Alsing$^1$}
\date{\normalsize%
    $^1$Air Force Research Laboratory, Rome, NY\\%
    $^2$Griffiss Institute, Rome, NY\\%
    $^3$EECS, Syracuse University, Syracuse, NY\\[2ex]%
    \today
}

\begin{document} 
\pagestyle{plain} 
\setcounter{page}{1} 

\maketitle

\centerline{\large\bf Abstract}
\skp{-0.4}
{\small
\begin{quote}
In this work we study 
tensor product operators on a cartesian product of finite dimensional Hilbert spaces into a target Hilbert space spanned by their image. 
A tensor product operator on a pair of Hilbert spaces is a maximally general bilinear operator into a target Hilbert space.  By \textit{maximally general} here, it is meant that every bilinear operator from the same pair of spaces to any Hilbert space factors into the composition of the tensor product operator with a uniquely determined linear mapping on the target space.   
There are multiple distinct tensor product operators of the same \textit{type}.  Distinctly different tensor product operators can be associated with different parts of a multipartite system and can be taken into account when combining states of the parts into a state of the whole.   Separability of states, and locality of operators and observables is tensor product operator dependent. Specifically, the same state in the target state space can be inseparable with respect to one tensor product operator but separable with respect to another, and no tensor product operator is distinguished relative to the others.  Since tensor product operators in multipartite systems are not usually specified, they are effectively determined by the specific choice of linear operators used to serve as unitary operations or as observables.
The unitary operator used to construct a Bell state from a pair of $\ket{0}$'s being highly tensor product operator-dependent is a prime example.  The relationship between two tensor product operators of the same type is given by composition with a unitary operator.  There is an equivalence between change of tensor product operator and change of basis in the target space of the tensor product operator.  For any effect that can be achieved with a change of tensor product operator together with a unitary operator or a projective measurement, that effect can be achieved by a change of basis in the target space together with the operator, and conversely.   Among the effects that can be achieved is the localization of some nonlocal operators as well as separability of inseparable states. 
In this work, we present and explore examples of these efffects.
\end{quote}
}

\section{Introduction}
{\small
\begin{quote}
Postulate 4: The state space of a composite physical system is \textit{the} [italics ours] tensor product
of the state spaces of the component physical systems. Moreover, if we have
systems numbered $1$ through $n$, and system number $i$ is prepared in the state
$\ket{\psi_i}$, then the joint state of the total system is $\ket{\psi_1}\otimes \ket{\psi_2}\otimes \cdots \otimes\ket{\psi_n}$.
\end{quote}

\skp{-1}
\hspace*{\fill} {\small Nielsen and Chuang, \textit{Quantum Computation and Information}, 2010, p.94}
}

There is no one, ``true" (i.e. uniquely preferred)  tensor product operator mapping a given pair of Hilbert spaces to a given target Hilbert space.    This is of course well known, but the details matter, and are not disposed of by appealing to isomorphism.  To do so is to ignore benefits that accrue from multiple distinct tensor product operators in the same multipartite system.  None of the operators are distinguished relative to the others any more than any orthonormal basis of a Hilbert space is distinguished relative to the others.   Separability of states and locality of unitary operators and observables is tensor product operator-dependent.  This statement stands in tension with the statement that
entanglement is physical.  In this introduction to the use of multiple distinct tensor product operators in multipartite quantum systems the examples that are given focus on (1) \textit{modularity} and (2) factorizing of states and operators.

A change of tensor product operator in a quantum program is a quantum program transformation.  
Work on quantum program verification, validation and optimization needs an underlying machine independent mathematically rigorous semantics, \cite{Ab04}, \cite{AJ94}, that should take account of a variety of program transformations.   
Due to modularity, 
the change of tensor product operator is among these transformations.  The term \textit{modularity} refers to the common-place practice that we build more complex systems out of more basic systems that have been previously constructed and supplied.  In the case of quantum systems it is easy to imagine a quantum circuit built from previously independently constructed circuits, each using potentially distinctly different tensor product operators. The details of combining these previously constructed circuits into a more complex circuit is described below.   In particular, change of tensor product operator for quantum circuits in ``mid-circuit'' can be effectively produced using swaps and, if necessary, similarity transformations.  An example is given below.

\subsection{Outline of the paper} 

Section~(\ref{sec:multiples}) consists of two subsections.  In subsection~(\ref{sse:bia2tpos}) a basis-independent construction of tensor product operators as maximally general bilinear operators is given following the procedures described in \cite{Ger85}.  In subsection~(\ref{sse:dependency}) the dependency of separability of states on the tensor product operator is shown.  Section~(\ref{se:unitary}) states and proves a lemma that precisely shows the unitary relationship between two distinct tensor product operators of the same type.  In that section a corollary of the lemma is stated and proved that gives an equivalence between change of tensor product operators and unitary similarity for general linear operators on the tensor product operator target space.  Section~(\ref{coord}) concentrates on coordinating unitary operators and measurements with tensor product operators. In subsection~(\ref{subcoord}) an example is given of a simple circuit to show the effect of a change of the tensor product operator based on the example of subsection~(\ref{sse:dependency}).
The two versions of the circuit are combined to show a how a change of tensor product operator may be uniformly implemented in
 ``mid-circuit,'' as well as to give an example of a localization of a $2$-qubit gate.   Subsection~(\ref{subsec:mixing}) presents a standard example of a teleportation circuit from
\cite{NC10} but in which distinct tensor products among the three pairs of qubits are used and combined.  The purpose is not to show there is a gain but that the modularity presents no difficulty and does not the change the physics.  We conclude with a brief section on future work to be undertaken.

\section{\large Multiple tensor product operators}\label{sec:multiples}
\subsection{A basis-independent approach to tensor product operators}\label{sse:bia2tpos}
We give here the definition of a tensor product operator as a maximally general bilinear operator.  The focus is on bilinear rather than multilinear tensor product operators from which multilinear operators are derivable. The definition is mildly category-theoretic without the diagrams, is basis independent, and follows \cite{Ger85}.  We begin by mentioning some convenient terminology regarding relaxed but still common and rigorous use of the term \textit{type} in its elementary sense to be used here and in the rest of the paper.

We are throughout this paper concerned with exploiting the multiple tensor product operators available for each \textit{type}.  The \textit{type} of a function $f$ mapping a set $S_1$ to a set $S_2$, denoted by $f\!:S_1 \longrightarrow S_2$ is the ordered pair of sets $(S_1, S_2)$ and is denoted $S_1 \longrightarrow S_2$.  $S_1$ is the \textit{domain} of $f$ and $S_2$ is the 
\textit {codomain} of $f$. They are components of $f$;  if either the domain or codomain is altered, the function is altered.  If the domain and codomain of a function are spaces, such as Hilbert spaces as they are here, the codomain is also called the \textit{target space}. 

A bilinear function $\otimes: \mathcal{V}_1 \cross \mathcal{V}_2 \longrightarrow \mathcal{V}$ is a tensor product operator if all bilinear operators defined on $\mathcal{V}_1 \cross \mathcal{V}_2$ uniquely factorize into a linear operator and $\otimes$.  Specifically,  for every bilinear operator $\beta: \mathcal{V}_1 \cross \mathcal{V}_2 \longrightarrow \mathcal{V}^\prime$, there is a \textit{unique} linear function $L: \mathcal{V} \longrightarrow \mathcal{V}^\prime$ such that $\beta = L\circ\otimes$; i.e. for all $(\bm{x},\bm{y}) \in \mathcal{V}_1 \cross \mathcal{V}_2$, $\beta(\bm{x},\bm{y}) = L(\bm{x} \otimes \bm{y})$.  This makes a tensor product operator $\otimes$ defined on $\mathcal{V}_1 \cross \mathcal{V}_2$ a maximally general bilinear operator on $\mathcal{V}_1 \cross \mathcal{V}_2$.  For Hilbert spaces, a tensor product operation is also required to preserve norms: if $\|\ket{x}\| = \|\ket{y}\| = 1$, then $\|\ket{x} \otimes\ket{y}\| = 1$.   

If $\otimes_1,\,\otimes_2 : \mathcal{V}_1 \cross \mathcal{V}_2 \longrightarrow \mathcal{V}$ are both tensor product operators, then both are bilinear and hence there are unique linear operators $L_1$ and $L_2$ such that $\otimes_2= L_1\circ\otimes_1$ and $\otimes_1= L_2\circ\otimes_2$.  Also there is are unique linear operators $I_1$ and $I_2$ such that $\otimes_1= I_1\circ\otimes_1$ and $\otimes_2= I_2\circ\otimes_2$.  It follows that $I_1$ and $I_2$ are both the identity function, and
$I_1 = L_2\circ L_1$ and $I_2 = L_1\circ L_2$. Thus $L_1$ and $L_2$ inverse to each other.  In the case of Hilbert spaces, since $\otimes_1$ and $\otimes_2$ are norm-preserving, $L_1$ and $L_2$ are forced to be norm-preserving, and therefore unitary\label{pr:unitary}.  

At this point, all bilinear tensor product operators of each type have been defined and shown to be related in a basis-independent manner by composition with a unitary operator on the target space of the type.

\subsection{Tensor product operator-dependency}\label{sse:dependency}

In this subsection an example of two tensor product operators is given for which the Bell states corresponding to each of the tensor product operators are shown to be separable with respect to the other operator.
We begin by reviewing the ordered basis-dependent approach to constructing a tensor product operator of a given type and then give examples of tensor product operator-dependency of entanglement of pure states, unitary operators and observables.   

A particular tensor product operator $\otimes$ of type $\mathcal{H}_1 \cross \mathcal{H}_2 \longrightarrow \mathcal{H}$ where $\mathcal{H}_1$, $\mathcal{H}_2$ and $\mathcal{H}$ are Hilbert spaces can be constructed by choosing ordered orthonormal bases $\bm{B}(\mathcal{H}_1)$, $\bm{B}(\mathcal{H}_2)$ and $\bm{B}(\mathcal{H})$ for these spaces, and choosing any \textit{bijective} order-preserving function $\gamma\!: \bm{B}(\mathcal{H}_1) \cross \bm{B}(\mathcal{H}_2) \longrightarrow \bm{B}(\mathcal{H})$. For this purpose, the order on $\mathcal{H}_1 \cross \mathcal{H}_2$ is lexicographically induced by the orders on $\mathcal{H}_1$ and $\mathcal{H}_2$.  That $\gamma$ is a bijection constrains the dimension of $\mathcal{H}$ in relation to the dimensions of $\mathcal{H}_1$ and $\mathcal{H}_2$.  It is then easy to prove by using the chosen bases that the uniquely determined bilinear extension of $\gamma$ to the whole of $\mathcal{H}_1 \cross \mathcal{H}_2$ satisfies the basis-independent requirement for being a tensor product operator.   For fixed bases of $\mathcal{H}_1$ and 
$\mathcal{H}_2$, distinct ordered orthonormal bases in $\mathcal{H}$ are in one-to-one correspondence with distinct tensor product operators of type $\mathcal{H}_1 \cross \mathcal{H}_2 \longrightarrow \mathcal{H}$.

As mentioned above, separability of states is tensor product \textit{operator-dependent}.   None of the  multiple distinct tensor product operators of the same \textit{type} are distinguished relative to the others any more than any of the orthonormal bases for a vector space are distinguished relative to the others.  

\begin{Example}\follows\label{ex:2tnsrs} {\em This is an example of two tensor product operators 
$\otimes_1$ and $\otimes_2$ with unequal corresponding Bell states, each of which is separable with respect to the other's associated tensor product operator.  

Take $\mathbb{C}^2$ with the usual ordered orthonormal basis, which we denote by $\bm{B}_0$.  $\bm{B}_0 = \{\ket{0} = (1,\,0),\ \ket{1} = (0,\,1)\}$, and
take $\mathbb{C}^4$ as the tensor product space with basis $\bm{B}_1$ to be the usual ordered orthonormal basis.  

Let $\bm{B}_2$ be the re-ordering of $\bm{B}_1$ obtained by applying the non-local controlled-not operator $CX$, to the basis vectors in $\bm{B}_1$. 

The column vectors and matrices in this example carry as subscripts the bases with respect to which they are written.  $\bm{B}_1$ is then given by,
\begin{equation}\label{eq:B1}
\bm{B}_1 = \left\{
{\left[\begin{matrix} 
1 \\ 0\\ 0\\ 0
\end{matrix}\right]\!\!,}_{\!\!\!\bm{B}_1}\ %
{\left[\begin{matrix} 
0 \\ 1\\ 0\\ 0
\end{matrix}\right]\!\!,}_{\!\!\!\bm{B}_1}\ %
{\left[\begin{matrix} 
0 \\ 0\\ 1\\ 0
\end{matrix}\right]\!\!,}_{\!\!\!\bm{B}_1}\ %
{\left[\begin{matrix} 
0 \\ 0\\ 0\\ 1
\end{matrix}\right]}_{\!\!\!\bm{B}_1}
\right\}.
\end{equation}
$\bm{B}_2$ is a permutation of $\bm{B}_1$:
\begin{equation}\label{eq:B2}
\bm{B}_2 = \left\{
{\left[\begin{matrix} 
1 \\ 0\\ 0\\ 0
\end{matrix}\right]\!\!,}_{\!\!\!\bm{B}_1}\ %
{\left[\begin{matrix} 
0 \\ 1\\ 0\\ 0
\end{matrix}\right]\!\!,}_{\!\!\!\bm{B}_1}\ %
{\left[\begin{matrix} 
0 \\ 0\\ 0\\ 1
\end{matrix}\right]\!\!,}_{\!\!\!\bm{B}_1}\ %
{\left[\begin{matrix} 
0 \\ 0\\ 1\\ 0
\end{matrix}\right]\!\!}_{\,\bm{B}_1}
\right\} = 
\left\{
{\left[\begin{matrix} 
1 \\ 0\\ 0\\ 0
\end{matrix}\right]\!\!,}_{\!\!\!\bm{B}_2}\ %
{\left[\begin{matrix} 
0 \\ 1\\ 0\\ 0
\end{matrix}\right]\!\!,}_{\!\!\!\bm{B}_2}\ %
{\left[\begin{matrix} 
0 \\ 0\\ 1\\ 0
\end{matrix}\right]\!\!,}_{\!\!\!\bm{B}_2}\ %
{\left[\begin{matrix} 
0 \\ 0\\ 0\\ 1
\end{matrix}\right]}_{\!\!\bm{B}_2}
\right\}.
\end{equation}
A controlled-not operator is clearly dependent on the computational basis in hand.  In particular, $CX$ relative to ${\bm{B}_1}$ is given by the familiar self-adjoint Hermitian matrix, 
\begin{equation}\label{eq:W}
CX = 
\left[\begin{matrix}
1 & 0 & 0 & 0 \\
0 & 1 & 0 & 0 \\
0 & 0 & 0 & 1 \\
0 & 0 & 1 & 0
\end{matrix}\right]_{\bm{B}_1} \!\!\!\!\!= 
\left[\begin{matrix}
1 & 0 & 0 & 0 \\
0 & 1 & 0 & 0 \\
0 & 0 & 0 & 1 \\
0 & 0 & 1 & 0
\end{matrix}\right]_{\bm{B}_2}.
\end{equation}

\skp{-0.4}
Since $CX$ is self-adjoint, the matrix for $CX$ has the same form relative to both bases in this case.  (Being self-adjoint is not essential to this and subsequent examples, except where specifically noted.)  Let $\otimes_2 = CX \circ \otimes_1$, and therefore $\otimes_1 = (CX)^\dagger \circ \otimes_2 = CX \circ \otimes_2$.  In the following equation, (\ref{eq:tnsrDpndntEntnglmnt}), $CX$ is non-local, so the result is not surprising.  
However, neither of these two tensor product operators is distinguished relative to the other, despite one of the ordered bases, 
$\bm{B}_1$, being standard, the other not.  (Just change the coordinate system.)
Then, 

\skp{-1}
\begin{footnotesize}
\begin{equation}\label{eq:tnsrDpndntEntnglmnt}\textstyle
\begin{array}{c}
\ket{\beta_{00}}_{\otimes_1} = \frac{1}{\sqrt{2}}(\ket{0}\otimes_{1}\ket{0} + \ket{1}\otimes_{1}\ket{1}) = CX\!\left(\frac{1}{\sqrt{2}} (\ket{0}\otimes_{2}\ket{0} + \ket{1}\otimes_{2}\ket{1})\right) = \ket{+}\otimes_{2}\ket{0}, \\
\rule{0pt}{3ex}
\ket{\beta_{00}}_{\otimes_2} = \frac{1}{\sqrt{2}}(\ket{0}\otimes_{2}\ket{0} + \ket{1}\otimes_{2}\ket{1}) = CX\!\left(\frac{1}{\sqrt{2}}(\ket{0}\otimes_{1}\ket{0} + \ket{1}\otimes_{1}\ket{1})\right) = \ket{+}\otimes_{1}\ket{0}.
\end{array}
\end{equation}
\end{footnotesize}

\skp{-1}
The equations in~(\ref{eq:tnsrDpndntEntnglmnt}) 
show that the indicated Bell state corresponding to each of the tensor product operators, 
$\otimes_{1}$, $\otimes_{2}$, is separable relative to the other. 

Note that the expected basic algebraic rules for Kronecker product operators on matrices holds, provided the matrices are written with respect to the appropriate ordered bases.

Restricting to a standard or preferred basis and preferred corresponding tensor product operator 
ignores the issue of what benefits may accrue from multiple tensor product operations in the same multipartite system.  A better approach is to \textit{coordinate} observables and unitary operators to tensor product operators, as we make clear in the examples in section~(\ref{coord}) below.   

In this subsection 
we have given an explicit example 
of two tensor product operators for which Bell states corresponding to each of the tensor product operators were shown to be separable with respect to the other operator. 
}
\end{Example}

\section{\large The unitary relationship between tensor product operators}\label{se:unitary}

\skp{-1}
In this section we examine
the relationship between two tensor product operators.  
Within a type $\mathcal{H}_1 \cross \mathcal{H}_2 \longrightarrow \mathcal{H}$, the relationship between two tensor product operators $\otimes_1$ and $\otimes_2$ is given by a uniquely determined unitary operator $U_{\otimes_1,\otimes_2}$ on the target space of the type.  $U_{\otimes_1,\otimes_2}$ is given in terms of the action of the two tensor product operators on the bases of $\mathcal{H}_1$ and $\mathcal{H}_2$.

The effect on vectors and linear transformations due to a change of tensor product operator is given by the following lemma and its proof.
Additionally, we prove a corollary that exhibits an equivalence between (1) change of tensor product operators and (2) unitary similarity of linear operators on the target space of the tensor product operators. 

\begin{Lemma}\follows\label{le:lem1} {\em 
For $i = a,b$, let $\otimes_{i}\!: \mathcal{H}_1 \cross \mathcal{H}_2 \longrightarrow \mathcal{H}$ be two tensor product operators, and for $j = 1,2$, let $\bm{B}(\mathcal{H}_j)$ be any orthonormal basis of $\mathcal{H}_j$.  Then

[Vectors] for a uniquely determined unitary operator $W$ on $\mathcal{H}$,
\begin{equation}\label{lem1.1}
\otimes_b = W\circ \otimes_a\,,
\end{equation}
[Vectors] $W,$ the unitary operator in~(\ref{lem1.1}), is uniquely determined by satisfying:
\begin{equation}\label{lem1.2}
W(\ket{e_i}\otimes_a\ket{f_j}) = \ket{e_i}\otimes_b\ket{f_j},\ \ \mbox{for all}\ (\ket{e_i},\ket{f_j}) \in \bm{B}(\mathcal{H}_1) \cross \bm{B}(\mathcal{H}_2),
\end{equation}
[Operators] If $L_i$, $i = 1,2,$ are any two linear operators on $\mathcal{H}_i$, $i=1,2$, respectively, and $W$ is the operator in ~(\ref{lem1.1}), then $L_1 \otimes_a L_2$ and $L_1 \otimes_b L_2$ are unitarily equivalent via $W$:
\begin{equation}\label{lem1.3}
L_1 \otimes_b L_2 = W(L_1 \otimes_a L_2)W^\dagger.
\end{equation}
}
\end{Lemma} 
\textit{Proof:} The existence and uniqueness of a unitary operator $W$ satisfying~(\ref{lem1.1}) has been proved in the argument of subsection~(\ref{sse:bia2tpos}).   
Equation~(\ref{lem1.2}), explicitly useful for matrix calculations, is a trivial consequence of~(\ref{lem1.1}).  For~(\ref{lem1.3}) we have,
\[
\begin{array}{l}
(L_1 \otimes_b L_2)(\ket{x}\otimes_b \ket{y})
= L_1\ket{x} \otimes_b L_2\ket{y}
= W(L_1\ket{x} \otimes_a L_2\ket{y})\\
\rule{0pt}{3ex} \!
= W(L_1\otimes_a L_2)(\ket{x}\otimes_a \ket{y})
= W(L_1\otimes_a L_2)W^\dagger(\ket{x}\otimes_b \ket{y}).
\end{array}
\]
\hspace*{\fill}\QED

For any bipartite quantum circuit \textbf{\sf C} using $\otimes$, consider the circuit \textbf{\sf C}$^\prime$ that results from 
\textbf{\sf C} obtained by replacing $\otimes$ by $\otimes^\prime$ and every gate and observable by the result of the similarity transform $W\rule{1.5ex}{0.7pt}\,W^\dagger$.   The corollary below shows that
$\textbf{\sf C}^\prime$ is equivalent to \textbf{\sf C} in the sense that the evolution of any state $W\ket{\psi_1}$ through \textbf{\sf C}$^\prime$ to a state $W\ket{\psi_2}$ corresponds to the evolution of $\ket{\psi_1}$ through \textbf{\sf C} to $\ket{\psi_2}$.   This is shown by~(\ref{cor.2}) below.  

\begin{Corollary}\follows\label{cor:lem1} {\em
Suppose $\otimes,\,\otimes^\prime\!: \mathcal{H}_1 \cross \mathcal{H}_2 \longrightarrow \mathcal{H}$ are tensor product operators with $\otimes^\prime = W\circ \otimes$, 
$L$ is a linear operator on $\mathcal{H}$,  
and $L_1$ and $L_2$ are linear operators on $\mathcal{H}_1$ and $\mathcal{H}_2$, respectively.  Then
\begin{equation}\label{cor.1}
W^\dagger L W = L_1 \otimes L_2\ \ \mbox{if, and only if,}\ \ L = L_1 \otimes^\prime L_2\,.
\end{equation}
and for any vector $\ket{x} \in \mathcal{H}$
\begin{equation}\label{cor.2}
(WLW^\dagger)W\ket{x} = WL\ket{x}.
\end{equation}
}
\end{Corollary}
\textit{Proof}: %
For~(\ref{cor.1}) we have,
\begin{equation}
W^\dagger L W = L_1 \otimes L_2\ \ \mbox{iff}\ \ L = W(L_1 \otimes L_2)W^\dagger = L_1 \otimes^\prime L_2\,.
\end{equation}
where the second equality follows from~(\ref{lem1.3}) in the lemma.   (\ref{cor.2}) is trivial, but nevertheless important as remarked upon above.  
\hspace*{\fill}\QED


The unitary relationship between tensor product operators and the effect on vectors and linear transformations due to a change of tensor product operator has been made explicit by the above lemma and its corollary.
These results reveal the equivalence between change of tensor product operators and unitary similarity of linear operators and observables.

\section{\large Coordination of gates, measurements and tensor product operators}\label{coord}


In this section we present examples showing how quantum gates and measurements are affected by several different tensor product operators applied to different pairs of qubits in the same circuit.  In particular, basis-dependent gates such as $CX$ and observables need to be \textit{coordinated}, in a manner to be made precise, with tensor product operators, since  separability of states and operators is tensor product operator-dependent.

With respect to a single tensor product operator, any general linear operator may fail to be local, yet \textit{may be} unitarily similar to a local operator.   The term \textit{localizable operator} is used in that case.

There exists  operators, 
e.g. $CX$ and $CZ$, that are nonlocalizable
relative to any ordered orthonormal basis are said to be \textit{non-localazable}.  Since $CX$ and $CZ$ are self-adjoint they also serve as nonlocalizable observables.  
{\small
\subsection{An example of coordination}\label{subcoord}
}
In this subsection we present a standard example of a teleportation circuit from
\cite{NC10} but in which distinct tensor products among the three pairs of qubits are used and combined.  The purpose is not to show there is a gain but that the modularity presents no difficulty and does not the change the physics (as expected).   A nontrivial example of a localizable non-local gate is given in this subsection.   

The formation of a Bell state, such as $\ket{\beta_{00}}$, is often given by applying a controlled-not operator to the result of applying a Hadamard operator to one of a pair of qubits initially in state 
$\ket{0} \otimes \ket{0}$:
\begin{equation}
(CX)(H\otimes I)(\ket{0}\otimes\ket{0}) = (CX)(H\ket{0} \otimes \ket{0})
= \frac{1}{\sqrt{2}}(\ket{0}\otimes\ket{0} + \ket{1}\otimes\ket{1})\,.
\end{equation}
In the example below we will use the two tensor product operators $\otimes_1$ and $\otimes_2$ from (\ref{eq:tnsrDpndntEntnglmnt}).    If $\otimes_1$ is the tensor product operator for the pair of qubits 
in the circuit below, then of course the Bell state $\ket{\beta_{00}}_{\otimes_1}$ is formed via the standard gate sequence 
(see, e.g., p27, \cite{NC10})
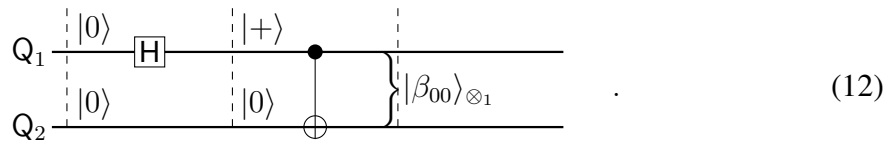
\begin{equation}\label{ct0a}
\raisebox{-3.84ex}{\mbox{\begin{tikzpicture}[scale=1.0]
\draw[line width=0.3mm] (0,0) -- (6.8,0);  
\draw[line width=0.3mm] (0,1) -- (1.1,1);
\draw[line width=0.3mm] (1.5,1) -- (6.8,1);


\draw[dashed] (0.2,-0.0) -- (0.2,1.6);
\node at (-0.3,1) {${\sf Q}_1$};
\node at (-0.3,0) {${\sf Q}_2$};
\node at (0.565,1.3) {$\ket{0}$};
\node at (0.565,0.3) {$\ket{0}$};

\draw[] (1.1,0.8) rectangle (1.5,1.2);
\node at (1.3,1.0) {\small\sf H};

\draw[dashed] (2.4,-0.0) -- (2.4,1.6);
\node at (2.804,1.3) {$\ket{+}$};
\node at (2.745,0.3) {$\ket{0}$};

\draw[] (3.5,-0.16) -- (3.5,1);
\fill[black] (3.5,1) circle (1mm);
\draw (3.5,0) circle (1.5mm);

\draw[dashed] (4.6,-0.0) -- (4.6,1.6);
\node at (5.1,0.5) {$\left. \rule{0pt}{3ex} \right\}\!\ket{\beta_{00}}_{\otimes_1}$};









\end{tikzpicture}}}\ \ \ \ \ \ .
\end{equation}
If for the pair of qubits $({\sf Q}_1,\,{\sf Q}_2)$, the operator $\otimes_1$ is replaced by $\otimes_2$,
then the Bell state $\ket{\beta_{00}}_{\otimes_2}$ is formed.   By~(\ref{eq:tnsrDpndntEntnglmnt}),
the state $\ket{+} \otimes_1 \ket{0}$ is also formed because it is identical to $\ket{\beta_{00}}_{\otimes_2}$.   Similarly, if in addition to replacing $\otimes_1$ by $\otimes_2$, the $CX$ gate is eliminated, then the $CX$ gate is effectively built into the tensor product operator for the pair $({\sf Q}_1,\,{\sf Q}_2)$ and the state formed, $\ket{+} \otimes_2 \ket{0}$, is $\ket{\beta_{00}}_{\otimes_1}$
as given by the circuit below:
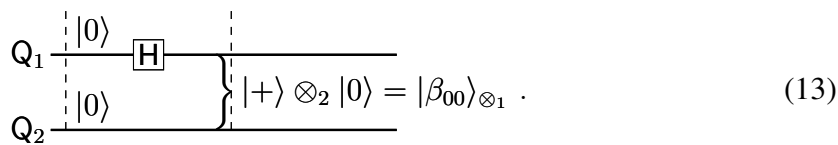
\begin{equation}\label{ct0b}
\raisebox{-3.84ex}{\mbox{\begin{tikzpicture}[scale=1.0]

\draw[line width=0.3mm] (0,0) -- (4.6,0);  

\draw[line width=0.3mm] (0,1) -- (1.1,1);
\draw[line width=0.3mm] (1.5,1) -- (4.6,1);



\draw[dashed] (0.2,-0.0) -- (0.2,1.6);
\node at (-0.3,1) {${\sf Q}_1$};
\node at (-0.3,0) {${\sf Q}_2$};
\node at (0.565,1.3) {$\ket{0}$};
\node at (0.565,0.3) {$\ket{0}$};

\draw[] (1.1,0.8) rectangle (1.5,1.2);
\node at (1.3,1.0) {\small\sf H};

\draw[dashed] (2.4,-0.0) -- (2.4,1.6);


\node at (4.095,0.5) {$\left. \rule{0pt}{3ex} \right\}\ket{+} \otimes_2 \ket{0} = \ket{\beta_{00}}_{\otimes_1}$};


\draw[line width=0.3mm] (0,0) -- (4.6,0);  

\draw[line width=0.3mm] (0,1) -- (1.1,1);
\draw[line width=0.3mm] (1.5,1) -- (4.6,1);



\draw[dashed] (0.2,-0.0) -- (0.2,1.6);
\node at (-0.3,1) {${\sf Q}_1$};
\node at (-0.3,0) {${\sf Q}_2$};
\node at (0.565,1.3) {$\ket{0}$};
\node at (0.565,0.3) {$\ket{0}$};

\draw[] (1.1,0.8) rectangle (1.5,1.2);
\node at (1.3,1.0) {\small\sf H};

\draw[dashed] (2.4,-0.0) -- (2.4,1.6);


\node at (4.095,0.5) {$\left. \rule{0pt}{3ex} \right\}\ket{+} \otimes_2 \ket{0} = \ket{\beta_{00}}_{\otimes_1}$};

\end{tikzpicture}}}.
\end{equation}

The point of this example is not to show there is a gain by eliminating the $CX$ gate. Rather, it is to show that \textit{both operators} (gates in the case of circuits) \textit{and observables} for multipartite systems must be chosen and coordinated with the tensor product operators that combine the states of the separate parts of such systems.

It might be argued that when we construct circuits we do not explicitly build in the tensor product operators among the qubits.  But 
in fact we do implicitly build in the tensor product operators by the choice of gates and observables in the circuit that are tensor product operator-dependent.  To change the tensor product operator in a circuit is to change the \textit{ordered} computational basis for the combined circuit, by re-ordering the computational basis, or changing the computational basis set altogether, or both. 

Suppose now that the state space of the circuit of both qubits is $\mathbb{C}^4$
with each of the 
qubits having state space $\mathbb{C}^2$.
In modifying~(\ref{ct0a}) to obtain~(\ref{ct0b}), the change from tensor product operator 
$\otimes_1$ to $\otimes_2$ 
re-orders the computational basis so that for example 
$\ket{1} \otimes_2 \ket{0}$ \textit{is} $\ket{1} \otimes_1 \ket{1}$   
and results in the overall state in circuit~(\ref{ct0b}) immediately after the Hadamard gate being the same as it is in~(\ref{ct0a}) after the $CX$ gate, as shown in the circuit diagram.  
Suppose ${\sf Q}_2$ is measured into its computational basis immediately following the $CX$ gate in (\ref{ct0a}) and the Hadamard gate in~(\ref{ct0b}).   In~(\ref{ct0a}) we obtain $\ket{0}_2$ or $\ket{1}_2$ with equal probability.
In~(\ref{ct0b}) we obtain $\ket{0}_2$ with certainty even though we are measuring the same state of the overall state of the circuit in each case.  We get different results because the (implicit) observables are \textit{different} in the two cases.  Specifically, 
an observable for the projective measurement into the computational basis of ${\sf Q}_2$, considered in isolation, is $Z$.
An observable $M$ for the measurement of the combined system of both qubits in~(\ref{ct0a}) is
given by $I \otimes_1 Z = I \otimes_1 (\ket{0}\bra{0} - \ket{1}\bra{1})$.   Let $P_1$ and $P_{-1}$ be the orthogonal projection operators in the spectral decomposition of $I \otimes_1 Z$.
In~(\ref{ct0b}), by lemma~(\ref{le:lem1}) the observable is given by 
$I \otimes_2 Z = (CX)(I \otimes_1 Z)(CX) = Z \otimes_1 Z = (\ket{0}\bra{0} - \ket{1}\bra{1})\otimes_1 (\ket{0}\bra{0} - \ket{1}\bra{1})$ with corresponding orthogonal projection operators $Q_1$ and $Q_2$.   Then for the circuit~(\ref{ct0a}) we have the two possible post-measurement states
\begin{equation}\label{obsTn1}
\frac{P_1\ket{\beta_{00}}_{\otimes_1}}{\sqrt{\bra{\beta_{00}}_{\otimes_1}P_{1}\ket{\beta_{00}}_{\otimes_1}}},\ \ \ \ \ \frac{P_{-1}\ket{\beta_{11}}_{\otimes_1}}{\sqrt{\bra{\beta_{11}}_{\otimes_1}P_{-1}\ket{\beta_{11}}_{\otimes_1}}}.
\end{equation}
In both cases the denominators, being $\frac{1}{2}$ show the post-measurement outcomes to be equally likely.   In~(\ref{ct0b}) the possible post-measurement states are
\begin{equation}\label{obsTn1}
\frac{Q_1\ket{\beta_{00}}_{\otimes_1}}{\sqrt{\bra{\beta_{00}}_{\otimes_1}Q_{1}\ket{\beta_{00}}_{\otimes_1}}},\ \ \ \ \ \frac{Q_{-1}\ket{\beta_{11}}_{\otimes_1}}{\sqrt{\bra{\beta_{11}}_{\otimes_1}Q_{-1}\ket{\beta_{11}}_{\otimes_1}}}.
\end{equation}

The second of these two post-measurement states is undefined, and the first has unit probably.
%
The distinct observables for these two circuits show that if the projective measurement into the computational basis of $\mathcal{H}({\sf Q}_1) \otimes \mathcal{H}({\sf Q}_2)$ as part~(\ref{ct0b}) is carried out with respect to the observable corresponding to $\otimes_2$ while $\otimes_{1}$ holds between ${\sf Q}_1$ and ${\sf Q}_2$, the improper coordination of the observable with the tensor product operator results in an unintended distribution on the possible results of the measurement.  

Change of tensor product operator from $\otimes_1$ to $\otimes_2$ in either of these circuits
can result in the localization of some measurements.
Consider a joint projective measurement of the qubits into the eigenstates of observable 
\begin{equation}
S =  \frac{1}{\sqrt{2}}
\left[\begin{array}{rrrr} 
1 &  \ \ 0 &  \ \ 0 &  -1 \\
0 & 1 &  -1 &  0 \\
0 &  -1 & -1 &  0 \\
-1 &  0 &  0 &  -1 
\end{array}\right]_{\bm{B}_1}
\end{equation}
which is a local with respect to $\otimes_2$ since 
\begin{equation}
S'=(CX)S(CX) = (\,\ket{-}\bra{0} - \ket{+}\bra{1}\,) \otimes_3 I\,.   
\end{equation}
The eigenvalues of $S$ and $S'$ are $1$ and $-1$, each with multiplicity $2$.    The corresponding measurement only nontrivially involves ${\sf Q}_1$ and measures ${\sf Q}_1$ into the eigenstates of $S'$. Thus, if the original eigenstates of $S$ are needed for further processing, they are known from the known observable and the post-measurement state that would have resulted from the original measurement.
Further, they recoverable from the known eigenstates of the observable. 

The relationship between $\otimes_1$ and $\otimes_2$ in circuit~(\ref{ct0a}) suggests that changing tensor products in ``mid-circuit'' can be useful.
Changing tensor product operators in this way is effectively achievable by keeping another related circuit but using a different tensor product operator in the other circuit.
One can then transfer
qubit states by swapping relative to the appropriate tensor product operators that connect the two circuits by connecting the corresponding qubits with swap gates.   For example, consider the following circuit
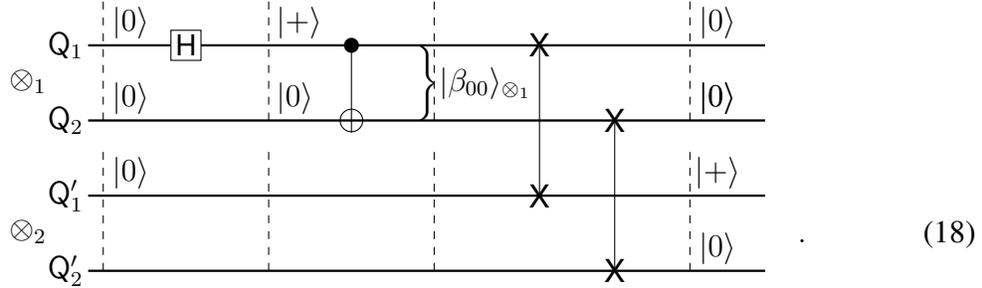
\begin{equation}\label{ct0c}
\raisebox{-3.84ex}{\mbox{\begin{tikzpicture}[scale=1.0]
\draw[line width=0.3mm] (0,2) -- (9,2);  
\draw[line width=0.3mm] (0,1) -- (9,1);  
\draw[line width=0.3mm] (0,3) -- (1.1,3);
\draw[line width=0.3mm] (1.5,3) -- (9,3);


\draw[dashed] (0.2,2) -- (0.2,3.6);
\node at (-0.3,3) {${\sf Q}_1$};
\node at (-0.3,2) {${\sf Q}_2$};
\node at (0.565,3.3) {$\ket{0}$};
\node at (0.565,2.3) {$\ket{0}$};
\node at (8.365,3.3) {$\ket{0}$};
\node at (8.365,2.3) {$\ket{0}$};
\node at (8.365,2.3) {$\ket{0}$};
\node at (8.365,1.3) {$\ket{+}$};
\node at (8.365,0.3) {$\ket{0}$};
\node at (-0.8,0.52) {$\otimes_2$};
\node at (-0.8,2.52) {$\otimes_1$};

\draw[] (1.1,2.8) rectangle (1.5,3.2);
\node at (1.3,3.0) {\small\sf H};

\draw[dashed] (2.4,2.0) -- (2.4,3.6);
\node at (2.804,3.3) {$\ket{+}$};
\node at (2.745,2.3) {$\ket{0}$};

\draw[] (3.5,1.84) -- (3.5,3);
\fill[black] (3.5,3) circle (1mm);
\draw (3.5,2) circle (1.5mm);

\draw[dashed] (4.6,2.0) -- (4.6,3.6);
\draw[dashed] (8,2.0) -- (8,3.6);
\node at (5.1,2.5) {$\left. \rule{0pt}{3ex} \right\}\!\ket{\beta_{00}}_{\otimes_1}$};









\draw[] (6,3) -- (6,1);
\draw[] (7,2) -- (7,0);
\node at (6,3) {\sf X};
\node at (6,1) {\sf X};
\node at (7,2) {\sf X};
\node at (7,0) {\sf X};


\draw[line width=0.3mm] (0,0) -- (9,0);  


\draw[dashed] (0.2,-0.0) -- (0.2,1.6);
\node at (-0.3,1) {${\sf Q}'_1$};
\node at (-0.3,0) {${\sf Q}'_2$};
\node at (0.565,1.3) {$\ket{0}$};


\draw[dashed] (2.4,-0.0) -- (2.4,1.6);



\draw[dashed] (4.6,-0.0) -- (4.6,1.6);
\draw[dashed] (8,-0.0) -- (8,1.6);









\end{tikzpicture}}} \quad.
\end{equation}
The tensor product operator to connect ${\sf Q}_1$ with ${\sf Q}^\prime_1$ and ${\sf Q}_2$ with ${\sf Q}^\prime_2$ can be any tensor product operator one likes and could even be distinct for each of the qubit pairs to be connected with swap gates.
In practice, that operator would be determined by what unitary operator one chooses to be the swap gates, but one might as well choose, say, $\otimes_1$.

In the following subsection, we show 
how to combine these extra tensor product operators with $\otimes_1$ and $\otimes_2$ to produce states of the $4$-qubit circuit.


{\small
\subsection{Mixing tensor products: example with teleportation}
\label{subsec:mixing}
}
This subsection presents a standard example of a teleportation circuit from
\cite{NC10} but in which distinct tensor products among the three pairs of qubits are used and combined.  
The purpose of this exercise is not to show there is a gain 
but rather to demonstrate that the modularity presents no difficulty and does not change the physics.    
In addition, a nontrivial example of a localizable non-local operator is given in this subsection.   

Consider the following well-known 3-qubit quantum circuit to teleport the state of one of Alice's qubits to Bob's qubit where Alice and Bob share the Bell state $\ket{\beta_{00}}$ at the start of the process, \cite{NC10}.  

\begin{equation}\label{ct0}
\raisebox{-6.4ex}{\mbox
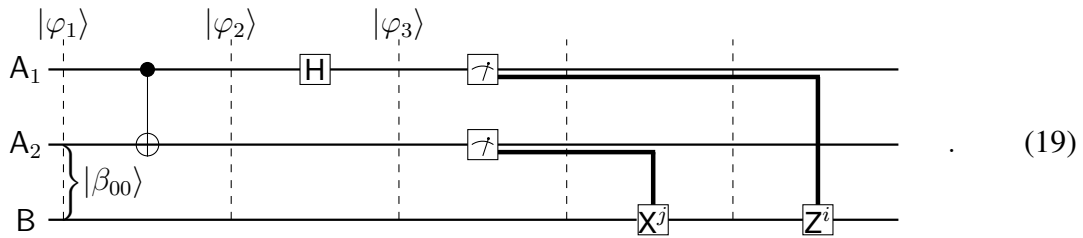
{\begin{tikzpicture}[scale=1.0]
\draw[dashed] (0.20,0) -- (0.20,2.4);               
\node at (0.20,2.6) {$\ket{\varphi_1}$};           
\draw[dashed] (2.43,0) -- (2.43,2.4);               
\node at (2.43,2.6) {$\ket{\varphi_2}$};           
\draw[dashed] (4.66,0) -- (4.66,2.4);               
\node at (4.66,2.6) {$\ket{\varphi_3}$};           
\draw[dashed] (6.89,0) -- (6.89,2.4);               
\draw[dashed] (9.10,0) -- (9.10,2.4);               

\node at (-0.3,2) {${\sf A}_1$};                              
\draw[line width=0.3mm] (0,2) -- (3.345,2);         
\fill[black] (1.317,2) circle (1mm);                        
\draw[] (3.345,1.8) rectangle (3.745,2.2);            
\node at (3.5448,2.0) {\small\sf H};                      
\draw[line width=0.3mm] (3.745,2) -- (5.575,2);  
\draw[] (5.575,1.8) rectangle (5.975,2.2);            
\draw[] (5.629,2) arc (125:55:0.25);                     
\draw[] (5.775,1.85) -- (5.85,2.08);                       
\fill[black] (5.85,2.08) circle (0.2mm);                   
\draw[line width=0.3mm] (5.975,2) -- (11.3,2);    
\draw[line width=0.6mm] (5.975,1.9) -- (10.257,1.9);    
\draw[line width=0.6mm] (10.23,0.2) -- (10.23,1.9); 

\node at (-0.3,1) {${\sf A}_2$};                             
\draw[line width=0.3mm] (0,1) -- (5.575,1);         
\draw (1.317,1) circle (1.5mm);                           
\draw[] (1.317,0.84) -- (1.317,2.0);                      
\draw[] (5.575,0.8) rectangle (5.975,1.2);            
\draw[] (5.629,1) arc (125:55:0.25);                     
\draw[] (5.775,0.85) -- (5.85,1.08);                      
\fill[black] (5.85,1.08) circle (0.2mm);                  
\draw[line width=0.3mm] (5.975,1) -- (11.3,1);    
\draw[line width=0.6mm] (5.975,0.9) -- (8.067,0.9);    
\draw[line width=0.6mm] (8.04,0.2) -- (8.04,0.9); 

\node at (-0.3,0) {${\sf B}$};                                
\draw[line width=0.3mm] (0,0) -- (7.82,0);          
\node at (0.69,0.5) {$\left. \rule{0pt}{3ex} \right\}\!\ket{\beta_{00}}$}; 
\draw[] (7.84,-0.2) rectangle (8.24,0.2);             
\node at (8.04,0.0) {{\small\sf X}$^{j}$};             
\draw[line width=0.3mm] (8.24,0) -- (10.03,0);   
\draw[] (10.03,-0.2) rectangle (10.43,0.2);          
\node at (10.23,0.0) {{\small\sf Z}$^{i}$};            
\draw[line width=0.3mm] (10.43,0) -- (11.3,0);    
\end{tikzpicture}}}\ \ \ \ \ \ .
\end{equation}

Below, we explore the consequences for this teleportation circuit arising from using distinct tensor product operators for each of the three 2-qubit subsystems.   The purpose is to exhibit in detail how potentially distinct tensor product operations among different pairs of qubits of a circuit affect the circuit's action. It will be seen that if the gates and observables on the $2$-qubit subsystems are coordinated with the tensor product operators on each of the $2$-qubit subsystems, then the net action of the circuit is to produce the same $3$-qubit states $\ket{\varphi_1}$, $\ket{\varphi_2}$ and $\ket{\varphi_3}$ as it would if it employed the same tensor product operator on all three $2$-qubit subsystems.

The state spaces of each of the three qubits is $\mathbb{C}^2$ with the usual standard ordered orthonormal basis which we have labeled $\bm{B}_0$, above.  For illustration purposes, we choose three distinct ordered orthonormal bases for the state spaces of the qubit pairs: $\mathcal{H}_{{\sf A}_1} \otimes \mathcal{H}_{{\sf A}_2}$, $\mathcal{H}_{{\sf A}_1} \otimes \mathcal{H}_{\sf B}$ and $\mathcal{H}_{{\sf A}_2} \otimes \mathcal{H}_{\sf B}$.  The tensor product symbol in the expressions for each of these \textit{state spaces for the qubit pairs} does not denote an operator on pairs of vectors. 
Instead, the symbol $\otimes$ denotes the tensor product space to serve as the codomain for tensor product operators on the vectors in the spaces being joined.
For example,
$\mathbb{C}^2 \otimes \mathbb{C}^2$ denotes the tensor product space that serves as the codomain for all tensor product operators of type $\mathbb{C}^2 \cross \mathbb{C}^2 \longrightarrow \mathbb{C}^2 \otimes \mathbb{C}^2$ such as the operators $\otimes_1$ and $\otimes_2$ that were defined above in example~(\ref{ex:2tnsrs}).   $\otimes_1$ and $\otimes_2$ have the same type: 
$\mathbb{C}^2 \cross \mathbb{C}^2 \longrightarrow \mathbb{C}^2 \otimes \mathbb{C}^2$, where it is assumed the tensor product space has been constructed to be $\mathbb{C}^4$.  

Two of the bases are given by:
$\bm{B}(\mathcal{H}_{{\sf A}_2} \otimes \mathcal{H}_{\sf B}) = \bm{B}_1$
and
$\bm{B}(\mathcal{H}_{{\sf A}_1} \otimes \mathcal{H}_{\sf B}) = \bm{B}_2$
and the third is given by
$\bm{B}(\mathcal{H}_{{\sf A}_1} \otimes \mathcal{H}_{{\sf A}_2}) = \bm{B}_3$,
where 
$\bm{B}_3$ is derived from the Bell basis with respect to $\otimes_1,\,$ and is given by
\begin{equation}
\begin{array}{rcrcr}
\ket{0}_{{\sf A}_1} \otimes_3 \ket{0}_{{\sf A}_2}\!\!\! & =\!\!\! & \ket{\beta_{10}}\!\!\! & =\!\!\! &
\frac{1}{\sqrt{2}}(\ket{0}_{{\sf A}_1} \otimes_1 \ket{0}_{{\sf A}_2} - \ket{1}_{{\sf A}_1} \otimes_1
\ket{1}_{{\sf A}_2}), \\
\rule{0pt}{2.6ex}\ket{0}_{{\sf A}_1} \otimes_3 \ket{1}_{{\sf A}_2}\!\!\! & =\!\!\! & \ket{\beta_{11}}\!\!\! & =\!\!\! & 
\frac{1}{\sqrt{2}}(\ket{0}_{{\sf A}_1} \otimes_1 \ket{1}_{{\sf A}_2} - \ket{1}_{{\sf A}_1} \otimes_1 \ket{0}_{{\sf A}_2}), \\
\rule{0pt}{2.6ex}\ket{1}_{{\sf A}_1} \otimes_3 \ket{0}_{{\sf A}_2}\!\!\! & =\!\!\! & -\ket{\beta_{01}}\!\!\! & =\!\!\! & 
\frac{-1}{\sqrt{2}}(\ket{0}_{{\sf A}_1} \otimes_1 \ket{1}_{{\sf A}_2} + \ket{1}_{{\sf A}_1} \otimes_1 \ket{0}_{{\sf A}_2}), \\
\rule{0pt}{2.6ex}\ket{1}_{{\sf A}_1} \otimes_3 \ket{1}_{{\sf A}_2}\!\!\! & =\!\!\! & -\ket{\beta_{00}}\!\!\! & =\!\!\! & 
\frac{-1}{\sqrt{2}}(\ket{0}_{{\sf A}_1} \otimes_1 \ket{0}_{{\sf A}_2} + \ket{1}_{{\sf A}_1} \otimes_1 \ket{1}_{{\sf A}_2}).
\end{array}
\end{equation}
Thus, $\otimes_3$ combines the subsystem consisting of qubits ${\sf A}_1$ and ${\sf A}_2$, $\otimes_2$ combines ${\sf A}_1$ and ${\sf B}$, and $\otimes_1$ combines ${\sf A}_2$ and ${\sf B}$.
The corresponding matrix for the similarity transformation $S^\dagger \underline{\ \ }\, S$ for a change of basis from $\bm{B}_1$ to $\bm{B}_3$ is
\begin{equation}\label{matrixS}
S = \frac{1}{\sqrt{2}}\left[
\begin{array}{rrrr}
       1 &        0 &        0 & \!\!\!-1 \\
       0 &        1 & \!\!\!-1 &        0 \\
       0 & \!\!\!-1 & \!\!\!-1 &        0 \\
\!\!\!-1 &        0 &        0 & \!\!\!-1
\end{array}
\right]_{\bm{B}_3 \rightarrow \bm{B}_1}\,.
\end{equation}


$\bm{B}_3$ yields a third tensor product operator $\otimes_3$ such that
\[
\bm{B}_3 = \{\ket{0}_{{\sf A}_1} \otimes_3 \ket{0}_{{\sf A}_2},\ \ket{0}_{{\sf A}_1} \otimes_3 \ket{1}_{{\sf A}_2},\ \ket{1}_{{\sf A}_1} \otimes_3 \ket{0}_{{\sf A}_2},\ \ket{1}_{{\sf A}_1} \otimes_3 \ket{1}_{{\sf A}_2}\}\,.
\]
Let $\mathcal{H} = \mathbb{C}^8$ be the state space of the circuit, and choose any fixed ordered orthonormal basis $\bm{B}$ of $\mathcal{H}$ and denote it by 
\begin{equation}\label{C^8Basis} 
\bm{B} = \{\ket{b_{ijk}} \mid i,j,k \in \{0,1\}\} =  \{\ket{b_{000}},  \ket{b_{001}}, \dots, \ket{b_{111}}\}.
\end{equation}
An expression such as
\begin{equation}\label{ill}
\ket{\psi_1}_{{\sf A}_1} \otimes_1 \ket{\psi_2}_{{\sf A}_2} \otimes_1 \ket{\psi_3}_{{\sf B}}
\end{equation}
is not rigorously defined.
To be rigorously defined the operator 
$\otimes_1$ would have to map, in particular, a pair of states, one of which is $4$-level, and the other $2$-level, to an $8$-level state while also mapping a pair of $2$-level states to a $4$-level state. 
Thus, the expression is \textit{ill-typed} due to both an implicit overloading of the symbol 
$\otimes_1$ and an ambiguous grouping.  A rigorous means of treating this obstacle is to
define several tensor product operators as in the following.  Define three tensor product operators: For each $i,j,k \in \{0,1\}$,
\begin{equation}
\begin{array}{l}
\otimes_{{\sf B}({\sf A}_1{\sf A}_2)}\!:(\mathcal{H}_{{\sf A}_1} \otimes \mathcal{H}_{{\sf A}_2}) \cross \mathcal{H}_{{\sf B}}\longrightarrow \mathcal{H}\!: (\ket{i}_{{\sf A}_1} \otimes_3 \ket{j}_{{\sf A}_2}, \ket{k}_{\sf B}) \mapsto \ket{b_{ijk}}, \\
\rule{0pt}{2.6ex}\otimes_{{\sf A}_2({\sf A}_1{\sf B})}\!:(\mathcal{H}_{{\sf A}_1} \otimes \mathcal{H}_{\sf B}) \cross \mathcal{H}_{{\sf A}_2}\longrightarrow \mathcal{H}\!: (\ket{i}_{{\sf A}_1} \otimes_2 \ket{k}_{\sf B}, \ket{j}_{{\sf A}_2}) \mapsto \ket{b_{ijk}}, \\
\rule{0pt}{2.6ex}\otimes_{{\sf A}_1({\sf A}_2{\sf B})}\!:(\mathcal{H}_{{\sf A}_2} \otimes \mathcal{H}_{\sf B}) \cross \mathcal{H}_{{\sf A}_1}\longrightarrow \mathcal{H}\!: (\ket{j}_{{\sf A}_2} \otimes_1 \ket{k}_{\sf B}, \ket{i}_{{\sf A}_1}) \mapsto \ket{b_{ijk}}.
\end{array}
\end{equation}

Prefix notation is used in these auxiliary tensor product operator definitions since it is necessary in the case of $\otimes_{{\sf A}_2({\sf A}_1{\sf B})}$.
With these operators the successive states of~(\ref{ct0}) can be calculated.  Let the initial state of ${\sf A}_1$ be $a\ket{0} + b\ket{1}$.
We then have

\begin{footnotesize}
\begin{equation}\label{phi1}
\begin{array}{rl}
& \ket{\varphi_1} \\
= & \\
& \otimes_{{\sf A}_1({\sf A}_2{\sf B})}(\ket{\beta_{00}}_{\otimes_1},\,a\ket{0}_{{\sf A}_1} + b\ket{1}_{{\sf A}_1}) \\
= & \\
& \textstyle{
\frac{a}{\sqrt{2}}(\otimes_{{\sf A}_1({\sf A}_2{\sf B})}(\ket{0}_{{\sf A}_2} \otimes_1 \ket{0}_{B},\,\ket{0}_{{\sf A}_1}))
+ \frac{a}{\sqrt{2}}(\otimes_{{\sf A}_1({\sf A}_2{\sf B})}(\ket{1}_{{\sf A}_2} \otimes_1 \ket{1}_{B},\,\ket{0}_{{\sf A}_1}))
} \\
& \rule{0pt}{4ex}\textstyle{
+\ \frac{b}{\sqrt{2}}(\otimes_{{\sf A}_1({\sf A}_2{\sf B})}(\ket{0}_{{\sf A}_2} \otimes_1 \ket{0}_{B},\,\ket{1}_{{\sf A}_1}))
+ \frac{b}{\sqrt{2}}(\otimes_{{\sf A}_1({\sf A}_2{\sf B})}(\ket{1}_{{\sf A}_2} \otimes_1 \ket{1}_{B},\,\ket{1}_{{\sf A}_1}))
} \\
= & \\
& \textstyle{\frac{a}{\sqrt{2}}\ket{b_{000}} + \frac{a}{\sqrt{2}}\ket{b_{011}} 
+ \frac{b}{\sqrt{2}}\ket{b_{100}} + \frac{b}{\sqrt{2}}\ket{b_{111}}} \\
= & \\
& \textstyle{
\frac{a}{\sqrt{2}}(\otimes_{{\sf B}({\sf A}_1{\sf A}_2)}(\ket{0}_{{\sf A}_1} \otimes_3 \ket{0}_{{\sf A}_2},\,\ket{0}_{\sf B}))
+ \frac{a}{\sqrt{2}}(\otimes_{{\sf B}({\sf A}_1{\sf A}_2)}(\ket{0}_{{\sf A}_1} \otimes_3 \ket{1}_{{\sf A}_2},\,\ket{1}_{\sf B}))
} \\
& \rule{0pt}{4ex}\textstyle{
+\ \frac{b}{\sqrt{2}}(\otimes_{{\sf B}({\sf A}_1{\sf A}_2)}(\ket{1}_{{\sf A}_1} \otimes_3 \ket{0}_{{\sf A}_2},\,\ket{0}_{\sf B}))
+ \frac{b}{\sqrt{2}}(\otimes_{{\sf B}({\sf A}_1{\sf A}_2)}(\ket{1}_{{\sf A}_1} \otimes_3 \ket{1}_{{\sf A}_2},\,\ket{1}_{\sf B}))\ .
} 
\end{array}
\end{equation}
\end{footnotesize}
\begin{footnotesize}
\begin{equation}\label{phi2}
\begin{array}{rl}
& \ket{\varphi_2} \\
= & \\
& ((CX)_{\bm{B}_3}\otimes_{{\sf B}({\sf A}_1{\sf A}_2)} I_{\sf B})\ket{\varphi_1} \\
= & \\
& 
\textstyle{
\frac{a}{\sqrt{2}}(\otimes_{{\sf B}({\sf A}_1{\sf A}_2)}(\ket{0}_{{\sf A}_1} \otimes_3 \ket{0}_{{\sf A}_2},\,\ket{0}_{\sf B}))
+ \frac{a}{\sqrt{2}}(\otimes_{{\sf B}({\sf A}_1{\sf A}_2)3(12)}(\ket{0}_{{\sf A}_1} \otimes_3 \ket{1}_{{\sf A}_2},\,\ket{1}_{\sf B}))
} \\
& \rule{0pt}{4ex}\textstyle{
+\ \frac{b}{\sqrt{2}}(\otimes_{{\sf B}({\sf A}_1{\sf A}_2)}(\ket{1}_{{\sf A}_1} \otimes_3 \ket{1}_{{\sf A}_2},\,\ket{0}_{\sf B}))
+ \frac{b}{\sqrt{2}}(\otimes_{{\sf B}({\sf A}_1{\sf A}_2)}(\ket{1}_{{\sf A}_1} \otimes_3 \ket{0}_{{\sf A}_2},\,\ket{1}_{\sf B}))
} \\
= & \\
& \textstyle{\frac{a}{\sqrt{2}}\ket{b_{000}} + \frac{a}{\sqrt{2}}\ket{b_{011}} 
+ \frac{b}{\sqrt{2}}\ket{b_{110}} + \frac{b}{\sqrt{2}}\ket{b_{101}}} \\
= & \\
& \textstyle{
\frac{a}{\sqrt{2}}(\otimes_{{\sf A}_1({\sf A}_2{\sf B})}(\ket{0}_{{\sf A}_2} \otimes_1 \ket{0}_{B},\,\ket{0}_{{\sf A}_1}))
+ \frac{a}{\sqrt{2}}(\otimes_{{\sf A}_1({\sf A}_2{\sf B})}(\ket{1}_{{\sf A}_2} \otimes_1 \ket{1}_{B},\,\ket{0}_{{\sf A}_1}))
} \\
& \rule{0pt}{4ex}\textstyle{
+\ \frac{b}{\sqrt{2}}(\otimes_{{\sf A}_1({\sf A}_2{\sf B})}(\ket{1}_{{\sf A}_2} \otimes_1 \ket{0}_{B},\,\ket{1}_{{\sf A}_1}))
+ \frac{b}{\sqrt{2}}(\otimes_{{\sf A}_1({\sf A}_2{\sf B})}(\ket{0}_{{\sf A}_2} \otimes_1 \ket{1}_{B},\,\ket{1}_{{\sf A}_1}))\,.
}
\end{array}
\end{equation}
\end{footnotesize}

\begin{footnotesize}
\begin{equation}\label{phi3}
\begin{array}{rl}
& \ket{\varphi_3} \\
= & \\
& (H_{{\sf A}_1} \otimes_{{\sf A}_1({\sf A}_2{\sf B})} I_{{\sf A}_2B})\ket{\varphi_2} \\
= & \\
&  \textstyle{
\frac{a}{\sqrt{2}}(\otimes_{{\sf A}_1({\sf A}_2{\sf B})}(\ket{0}_{{\sf A}_2} \otimes_1 \ket{0}_{B},\,\ket{+}_{{\sf A}_1}))
+ \frac{a}{\sqrt{2}}(\otimes_{{\sf A}_1({\sf A}_2{\sf B})}(\ket{1}_{{\sf A}_2} \otimes_1 \ket{1}_{B},\,\ket{+}_{{\sf A}_1}))
} \\
& \rule{0pt}{4ex}\textstyle{
+\ \frac{b}{\sqrt{2}}(\otimes_{{\sf A}_1({\sf A}_2{\sf B})}(\ket{1}_{{\sf A}_2} \otimes_1 \ket{0}_{B},\,\ket{-}_{{\sf A}_1}))
+ \frac{b}{\sqrt{2}}(\otimes_{{\sf A}_1({\sf A}_2{\sf B})}(\ket{0}_{{\sf A}_2} \otimes_1 \ket{1}_{B},\,\ket{-}_{{\sf A}_1}))
} \\
= & \\
& \textstyle{
\frac{a}{2}\ket{b_{000}} + \frac{a}{2}\ket{b_{100}} + %
\frac{a}{2}\ket{b_{011}} + \frac{a}{2}\ket{b_{111}} + %
\frac{b}{2}\ket{b_{010}} - \frac{b}{2}\ket{b_{110}} + %
\frac{b}{2}\ket{b_{001}} - \frac{b}{2}\ket{b_{101}}} \\
= & \\
& \textstyle{
\ \ \ \ \frac{a}{2}(\otimes_{{\sf A}_2({\sf A}_1{\sf B})}(\ket{0}_{{\sf A}_1} \otimes_2 \ket{0}_{B},\,\ket{0}_{{\sf A}_2})) +
\frac{a}{2}(\otimes_{{\sf A}_2({\sf A}_1{\sf B})}(\ket{1}_{{\sf A}_1} \otimes_2 \ket{0}_{B},\,\ket{0}_{{\sf A}_2}))
} \\
& \rule{0pt}{4ex}\textstyle{
+\ \frac{a}{2}(\otimes_{{\sf A}_2({\sf A}_1{\sf B})}(\ket{0}_{{\sf A}_1} \otimes_2 \ket{1}_{B},\,\ket{1}_{{\sf A}_2})) +
\frac{a}{2}(\otimes_{{\sf A}_2({\sf A}_1{\sf B})}(\ket{1}_{{\sf A}_1} \otimes_2 \ket{1}_{B},\,\ket{1}_{{\sf A}_2}))
} \\
& \rule{0pt}{4ex}\textstyle{
+\ \frac{b}{2}(\otimes_{{\sf A}_2({\sf A}_1{\sf B})}(\ket{0}_{{\sf A}_1} \otimes_2 \ket{0}_{B},\,\ket{1}_{{\sf A}_2})) -
\frac{b}{2}(\otimes_{{\sf A}_2({\sf A}_1{\sf B})}(\ket{1}_{{\sf A}_1} \otimes_2 \ket{0}_{B},\,\ket{1}_{{\sf A}_2}))
} \\
& \rule{0pt}{4ex}\textstyle{
+\ \frac{b}{2}(\otimes_{{\sf A}_2({\sf A}_1{\sf B})}(\ket{0}_{{\sf A}_1} \otimes_2 \ket{1}_{B},\,\ket{0}_{{\sf A}_2})) -
\frac{b}{2}(\otimes_{{\sf A}_2({\sf A}_1{\sf B})}(\ket{1}_{{\sf A}_1} \otimes_2 \ket{1}_{B},\,\ket{0}_{{\sf A}_2}))\,.
}
\end{array}
\end{equation}
\end{footnotesize}

The observables for projectively measuring Alice's qubits are
\begin{equation}\label{Aobs}
{\sf A}_1\!\!:\ \otimes_{{\sf A}_1({\sf A}_2{\sf B})}(I_{(\mathcal{H}_{{\sf A}_2} \otimes \mathcal{H}_{\sf B})},Z_{(\mathcal{H}_{{\sf A}_1})}),
\ \ \ \ \ \ %
{\sf A}_2\!\!:\ \otimes_{{\sf A}_2({\sf A}_1{\sf B})}(I_{(\mathcal{H}_{{\sf A}_1} \otimes \mathcal{H}_{\sf B})},Z_{(\mathcal{H}_{{\sf A}_2})}).
\end{equation}
The subscripts on the linear operators $I$ and $Z$ indicate the spaces on which the operators are defined.  The action of $\otimes_{{\sf A}_2({\sf A}_1{\sf B})}(I_{(\mathcal{H}_{{\sf A}_1} \otimes \mathcal{H}_{\sf B})},Z_{(\mathcal{H}_{{\sf A}_2})})$, for example, is given by
\begin{equation}
(\ket{i}_{{\sf A}_1} \otimes_2 \ket{k}_{\sf B}, \ket{j}_{{\sf A}_2}) \mapsto 
\otimes_{{\sf A}_2({\sf A}_1{\sf B})}(\ket{i}_{{\sf A}_1} \otimes_2 \ket{k}_{\sf B}, (-1)^j\ket{j}_{{\sf A}_2}) \mapsto (-1)^j\ket{b_{ijk}}.  
\end{equation}
Note that the observable for ${\sf A}_1$ can be written with infix notation:
\begin{equation}\label{infixprefix}
Z_{(\mathcal{H}_{{\sf A}_1})} \otimes_{{\sf A}_1({\sf A}_2{\sf B})} I_{(\mathcal{H}_{{\sf A}_2} \otimes \mathcal{H}_{\sf B})}\,.
\end{equation}
The observable for ${\sf A}_2$ cannot be written with infix notation because the conventional procedure for writing the outcome of the Kronecker product does not result in the correct lexicographic ordering of the resulting basis.  The spectral decompositions of the observables are
\begin{equation}\label{Aobs}
\begin{array}{l}
{\sf A}_1\!\!:\ \otimes_{{\sf A}_1({\sf A}_2{\sf B})}(I_{(\mathcal{H}_{{\sf A}_2} \otimes \mathcal{H}_{\sf B})},1\cdot\ket{0}_{{\sf A}_1}\bra{0}_{{\sf A}_1} + (-1)\cdot\ket{1}_{{\sf A}_1}\bra{1}_{{\sf A}_1}), \\
\rule{0pt}{2.6ex}
{\sf A}_2\!\!:\ \otimes_{{\sf A}_2({\sf A}_1{\sf B})}(I_{(\mathcal{H}_{{\sf A}_1} \otimes \mathcal{H}_{\sf B})},1\cdot\ket{0}_{{\sf A}_2}\bra{0}_{{\sf A}_2} + (-1)\cdot\ket{1}_{{\sf A}_1}\bra{1}_{{\sf A}_1}).
\end{array}
\end{equation}
The four possible outcomes of the two measurements amount to what is standardly obtained for a post-measurement state of the three qubits:
\begin{equation}\label{outcome}
\begin{array}{l}
(\ket{0}_{{\sf A}_1} \otimes_3 \ket{0}_{{\sf A}_2}) \otimes_{{\sf B}({\sf A}_1{\sf A}_2)}(a\ket{0}_{\sf B} + b\ket{1}_{\sf B}), \\
(\ket{0}_{{\sf A}_1} \otimes_3 \ket{1}_{{\sf A}_2}) \otimes_{{\sf B}({\sf A}_1{\sf A}_2)}(a\ket{1}_{\sf B} + b\ket{0}_{\sf B}), \\
(\ket{1}_{{\sf A}_1} \otimes_3 \ket{0}_{{\sf A}_2}) \otimes_{{\sf B}({\sf A}_1{\sf A}_2)}(a\ket{0}_{\sf B} - b\ket{1}_{\sf B}), \\
(\ket{1}_{{\sf A}_1} \otimes_3 \ket{1}_{{\sf A}_2}) \otimes_{{\sf B}({\sf A}_1{\sf A}_2)}(a\ket{1}_{\sf B} - b\ket{0}_{\sf B}),
\end{array}
\end{equation}
which are exactly what was to be expected from standard teleportation circuit for qubit {\sf B}.

Reiterating, 
the purpose of the previous exercise has not been to show there is a gain from using the these distinct tensor product pairs but rather that the use of multiple tensor product operators does not present difficulty and does not change the physics.  The example has also shown in detail how to combine the separate tensor product operators by using auxiliary operators, and further, to remind us that even if the same operator, say $\otimes_1$, had been used for each of the qubit pairs, auxiliary operators would still have been required.

\skp{0.5}
\section{Conclusion and Future work}
\skp{-0.5}
This introduction to the use of multiple tensor product operators on the same multipartite quantum system has provided examples of the effects that can be produced by actions that include change of tensor product operators.  Changing tensor product operators in ``mid-circuit'' is effectively achievable by keeping a second related circuit 
which uses a different tensor product operator. 
One then transfers qubit states between the two circuits by swapping corresponding qubits
relative to the appropriate tensor product operators that connect the two circuits.
The apparatus for this procedure has been exhibited by the examples discussed  in section~(\ref{coord}).  The formal requirements that need to be satisfied when combining more than two parts into a multipartite system using bilinear 
(as distinguished from multilinear) tensor product operators, has also been exhibited in subsection~(\ref{subsec:mixing}).

In future work we plan to 
investigate the  consequences of the use of multiple tensor products for quantum programming methodology and quantum protocol formulation. 
In addition, we plan to explore 
the characterization of the degrees of entanglement, as well as
questions of quantum computational subrecursive complexity,  algorithmic complexity, and recursion-theoretic complexity.
 In this work we have emphasized the equivalence of the change of tensor product operators and similarity transformations. 
Based on this effort, and  prior work on the factorizing of separable states and localizable operators \cite{BA13,BJA21}, 
we plan to conduct detailed investigations into quantum program transformation for program verification, as well as into the similarity classes within the unitary groups, 
and the relationship between these classes and tensor product operators.

\skp{1}
\begin{flushleft}
\textbf{\Large Acknowledgements}
\end{flushleft}
HAB would like to acknowledge support for this work from the Air Force Research Laboratory, 
Information Directorate.
PMA would like to acknowledge support of this work from
the Air Force Office of Scientific Research (AFOSR).
Any opinions, findings and conclusions or recommendations
expressed in this material are those of the author(s) and do not
necessarily reflect the views of Air Force Research Laboratory.

\bibliographystyle{plain}
\bibliography{quantcomp}

\begin{thebibliography}{1}

\bibitem{Ab04}
S.~Abramsky.
\newblock A cook's tour of a simple quantum programming language.
\newblock In {\em 3rd International Symposium on Domain Theory}, 2004.

\bibitem{AJ94}
S.~Abramsky and A.~Jung.
\newblock Domain theory.
\newblock In S.~Abramsky, A.~Jung, and T.~Maibaum, editors, {\em Handbook of
  Logic in Computer Science, vol. 3}, pages 1--168. Clarendon Press, Oxford,
  1994.

\bibitem{BA13}
H.A. Blair and P.M. Alsing.
\newblock Handy elementary algebraic properties of the geometry of
  entanglement.
\newblock In {\em Proceedings of SPIE Conference on Quantum Information and
  Computation XI, Vol. 8749}, 2013.

\bibitem{BJA21}
H.A. Blair, H.S.Jacinto, and P.M. Alsing.
\newblock Technical note: Vector and matrix factorization.
\newblock 2021.

\bibitem{Ger85}
R.~Geroch.
\newblock {\em Mathematical Physics}.
\newblock University of Chicago, 1985.

\bibitem{NC10}
M.A. Nielsen and I.L. Chuang.
\newblock {\em Quantum Computation and Quantum Information}.
\newblock Cambridge University Press, 2010.

\end{thebibliography}

\end{document}